\documentstyle[12pt]{article}
\begin{document}

\title {$\tau - \mu - e$ Universality in $\tau$ Decays and
Constraints on the Slepton Masses.
 \thanks{Supported in part by the Polish Committee for Scientific
         Research under the grant 2 0165 91 01}}
\author{Piotr H. Chankowski
\thanks{On leave of absence from
Institute of Theoretical Physics, Warsaw University.}\\
Istituto Nazionale di Fisica Nucleare, Sezione di Padova\\
and Dipartimento di Fisica "Galileo Galilei"\\
via F. Marzolo 8, 35131 Padova, Italy,\\
Ralf Hempfling \\
DESY, Notkestr. 85, 22603 Hamburg\\
and\\
Stefan Pokorski$^{\dagger}$\\
Max-Planck-Institute f\"ur Physik\\
-Werner-Heisenberg-Institute-\\
P.O.Box 40 12 12, Munich, Fed. Rep. Germany}

\maketitle

\vspace{-15cm}
\begin{flushright}
DFPD 94/TH-27\\
MPI-Ph/94-24\\
DESY 94-077\\
\end{flushright}
\vspace{15cm}

\begin{abstract}
The leptonic $\tau$ decays are calculated at the 1-loop level
in the Minimal Supersymmetric Standard Model. The deviation from the
$\tau - \mu - e$ universality is studied as a function of the
supersymmetric parameters and discussed in the context of the expected
improvement of the experimental accuracy.
\end{abstract}

\newpage

Studying the $\tau - \mu - e$ universality in the leptonic $\tau$
decays is an interesting laboratory for search for physics beyond the
Standard Model.

In the Standard Model the $\tau$ decay partial width for the leptonic
modes is:
\begin{eqnarray}
\Gamma\left(\tau\rightarrow l\bar\nu_l\nu_{\tau}(\gamma)\right)
&=& {G^2_{F} m^5_{\tau}\over192\pi^3} f(m^2_l/m^2_{\tau})\times\\
  &&\left[1 + {3\over5}{m^2_{\tau}\over M^2_W}\right]
  \left[1 + {\alpha(m_{\tau})\over2\pi}\left({25\over4}-\pi^2\right)\right]
\nonumber
\end{eqnarray}
where $f(x) = 1 - 8x + 8x^3 - x^4 -12x^2\log x$ is the lepton mass
correction and the last two factors are corrections from the nonlocal
structure of the intermediate $W^{\pm}$ boson propagator and QED
radiative corrections respectively. The Fermi constant $G_F$ is
determined by the muon life-time
\begin{eqnarray}
 G_F\equiv G_{\mu}= (1.16637\pm 0.00002)\times10^{-5}{\rm GeV}^{-2}
\end{eqnarray}
and absorbs all the remaining electroweak radiative (loop) corrections:
\begin{eqnarray}
{G_{\mu}\over\sqrt2} =
{\pi\alpha\over2M^2_W\left(1-M^2_W/M^2_Z\right)}
{1\over 1-\Delta r\left(\alpha,M_W,M_Z,m_t,...\right)}
\end{eqnarray}
In the on-shell renormalization scheme
\begin{eqnarray}
\Delta r = - {\hat\Pi^T_{WW}\left(0\right)\over M^2_W} + \Delta r^{\prime}
\end{eqnarray}
where $\hat\Pi^T_{WW}$ is the renormalized $W^{\pm}$ boson
self-energy calculated at zero momentum (process independent "oblique"
correction) and $\Delta r^{\prime}$ includes box and vertex corrections
as well as the wave function renormalization factors for external
neutrinos \cite{HOL}. In the Standard Model the corrections
$\Delta r^{\prime}$ are universal for all the decays
$l\rightarrow l^{\prime}\nu_l\bar \nu_{l^{\prime}}$ and hence follows
the prediction (1) for the $\tau$ decays \cite{MASI}. Any experimental
deviation from these predictions would indicate the presence of physics
beyond the Standard Model
\footnote{Fermion mass dependence of the SM boxes and vertices as well as
          external momentum effects also give rise to small departure from
          strict universality of $\Delta r^{\prime}$. This departure is,
          however, negligible being of order ${\cal O}(\alpha m^2_l/M^2_W)$.}.

The purpose of this letter is to study the deviation from the $\tau-\mu-e$
universality in the leptonic $\tau$ decays in the framework of the
Minimal Supersymmetric Standard Model (MSSM) (the quark-lepton
non-universality in the MSSM has been studied in ref.\cite{barb}).
The main source of non$-$universal contributions would be the tree level
contribution from the charged Higgs boson (mass dependent couplings) and
different slepton masses of the $\tilde\mu$, $\tilde\tau$ and $\tilde e$
sleptons exchanged in the loops. Thus the leptonic $\tau$ decays
offer a unique possibility to establish some limits on the
intergeneration mass splitting for sleptons
in absence of the intergeneration mixing.
Those limits are
complementary to the limits on the intergeneration mixing in the
slepton mass matrix, which can be derived from the FCNC transitions
\cite{GM}.

With new, non$-$universal contributions the
$G_{\mu}$ $-$ muon Fermi decay constant in eq.(1) has to be replaced by
the process dependent constant
\begin{eqnarray}
  G_{\mu}\rightarrow G_{\tau, l}
\end{eqnarray}
with $ G_{\mu}\equiv G_{\mu, e}$ in the present notation.

A new physics contribution to $G_{l,l^{\prime}}$ can be classified as
corrections to the strength of the effective $(V-A)\times (V-A)$
four$-$fermion interaction and corrections with another Lorentz structure
of the effective four$-$fermion interaction.
Supersymmetric particle exchange in the loops contributes mainly to the former
whereas the tree level charged Higgs boson (and Goldstone boson) exchange
contributes to the latter.
Hence, the full contribution (say at 1-loop
accuracy) to the $G_{l,l^{\prime}}$ defined by eq.(1) can be written
as \cite{KP}:
\begin{eqnarray}
G^2_{l,l^{\prime}}= \tilde G^2_{l,l^{\prime}}\left(1+{1\over32} H^2\right)
\left[1-{m_{\mu}\over m_{\tau}}
{H\over\sqrt{2}\left(1+{1\over32}H^2\right)}\right]
\end{eqnarray}
where
\begin{eqnarray}
\tilde G_{l,l^{\prime}} H =
{g^2_2\over2}{m_l m_{l^{\prime}}\over M^2_W}
\left({\tan^2\beta\over M^2_{H^{\pm}}} + {1\over M_W^2}\right)
\end{eqnarray}
and $\tilde G_{l,l^{\prime}}$ is the one-loop corrected Fermi constant
in the absence of the tree level Higgs contribution. It can be
parameterized as:
\begin{eqnarray}
{\tilde G_{l,l^{\prime}}\over\sqrt2} =
{\pi\alpha\over2M^2_W\left(1-M^2_W/M^2_Z\right)}
{1\over 1-\Delta r}\left(1 + \Delta r^{\prime}_{l,l^{\prime}}\right)
\end{eqnarray}
where $\Delta r$ includes now all the SM corrections and the process
independent ("oblique") supersymmetric corrections and
$\Delta r^{\prime}_{l,l^{\prime}}$ contains only process dependent
supersymmetric one-loop corrections.

The deviations from the $\tau-\mu-e$  universality can be
conveniently discussed by studying the ratios $G_{\tau,e}/G_{\mu,e}$,
$G_{\tau,\mu}/G_{\mu,e}$ and $G_{\tau,\mu}/G_{\tau,e}$, given by
the ratios of the corresponding branching fractions.
With the highly accurate experimental result for the $G_{\mu,e}$,
the first two ratios are essentially a direct measure of
non-universality in the corresponding tau decays.
                                              When the statistical
error of future experiments will become
      negligible, the main problem for achieving maximum
precision will be to reduce the systematic errors. One may
expect that certain systematic errors will be cancelled
in the ratio        $G_{\tau,\mu}/G_{\tau,e}$.

In case the tree level Higgs exchange is negligible, i.e. for
$\tau\rightarrow e\nu_{\tau}\bar\nu_e$ and
$\mu\rightarrow e\nu_{\mu}\bar\nu_e$ (see below) we have
$G_{l,l^{\prime}}=\tilde G_{l,l^{\prime}}$ and
\begin{eqnarray}
G_{\tau, e}/G_{\mu,e} = 1 + \Delta r^{\prime}_{\tau,e}
                          - \Delta r^{\prime}_{\mu,e}
\end{eqnarray}
For ratios involving $G_{\tau,\mu}$ the complete eq.(6) has to be used
(in particular for large $\tan\beta$)
\footnote{In many presentations the deviation from universality are
          parameterized in terms of the "effective" coupling
          constants $G_{l,l^{\prime}} = g_l g_{l^{\prime}}$. However, as
          we see, the new physics contribution, in general, do not factorize
          and this parameterization is unnatural.}.

We shall now discuss in more detail the new (supersymmetric)
contributions and estimate their magnitude.
The corrections $\Delta r^{\prime}_{l,l^{\prime}}$ contain the
box, wave function renormalization and vertex contributions from the
supersymmetric particle exchanges. The recent complete calculation of the
$\mu$ decay in the  MSSM
\cite{MY1} can be easily extended to study $\tau$ decays. For the
details of the $G_{\mu}$ calculation we refer the reader to the ref.
\cite{MASI,HOL,MY1}. The extension
to the case of $\tau$ decays requires the inclusion of the Higgs boson
exchanges (neglected in the calculation of the $\mu$ decay width)
as for large $\tan\beta$, the ratio of the vacuum expectation
values of the two Higgs doublets, they can compete with the generic
SUSY contributions. We remind that in the MSSM there are five physical
Higgs bosons: one $CP$-odd $A^0$, two $CP$-even $H^0$ and $h^0$
and two charged $H^{\pm}$.
The rough estimate of various contributions to the $\tau\rightarrow
l\nu_{\tau}\bar\nu_l$ decay amplitude is as follows. The tree level
charged Higgs boson exchange is suppressed as compared to the
dominant tree level $W^{\pm}$ exchange by the factor
\begin{eqnarray}
\left({m_{\tau} m_l\over M^2_{H^{\pm}}}\right)\tan^2\beta
\end{eqnarray}
Similar diagram generated by the charged Goldstone boson is much smaller
(for $M_{H^{\pm}}\sim M_W$) since the couplings of the Goldstone bosons
do not have the $\tan\beta$ amplification. The dominant diagrams
containing one loop
corrections to the $\tau W^{\pm}\nu_{\tau}$ vertex have the following
suppression factors (compared to the dominant tree level graph):\\
\begin{itemize}
\item{}
      ${g^2_2\over16\pi^2}\left({m^2_{\tau}\over M^2_W}\right)\tan\beta
      \left({M^2_W\over M^2_{H^0(h^0)}}\right)$ $-$
      for loops with $H^0(h^0)W^{\pm}\tau$ and $H^0(h^0)G^{\pm}\tau$
      exchange,
\item{}
      ${g^2_2\over16\pi^2}\left({m^2_{\tau}\over M^2_W}\right)
      \tan^2\beta\left({M_W^2\over M^2_{A^0}}\right)$ $-$
      for loops with $H^0(h^0)H^{\pm}\tau$ or $A^0H^{\pm}\tau$
      exchange,
\item{}
      ${g^2_2\over16\pi^2}\left({M^2_W\over M^2_{SUSY}}\right)$ $-$
      for genuine SUSY loops generated by chargi\-nos/ne\-utra\-linos
      and sleptons.
\end{itemize}
\noindent Other possible vertex diagrams do not contain $\tan\beta$.
Similar estimates with $m_{\tau}\rightarrow m_l$ hold for the
$l W^{\pm}\nu_l$ vertex. For dominant box contributions we have
the following estimates:
\begin{itemize}
\item{}
      ${g^2_2\over16\pi^2}\left({m_{\tau} m_l\over M^2}\right)
      \tan^2\beta$ $-$
      (where $M$ is the maximal mass circulating in the loop)
      for boxes with $H^0(h^0)W^{\pm}$, $A^0W^{\pm}$
      or $Z^0H^{\pm}$ exchange,
\item{}
      ${g^2_2\over16\pi^2}\left({m_{\tau}m_l\over
      M_WM_{H^{\pm}}}\right)^2\tan^4\beta$
      $-$ for boxes with $H^0(h^0)H^{\pm}$ or $A^0H^{\pm}$ exchange,
\item{}
      ${g^2_2\over16\pi^2}\left({M^2_W\over M^2_{SUSY}}\right)$ $-$
      (where $M_{SUSY}$ is the maximal SUSY mass circulating in the loop)
      for genuine SUSY loops generated by char\-gi\-nos/neu\-tralinos
      and sleptons.
\end{itemize}

  From the above estimations, taking into account that
${g^2_2\over16\pi^2}\sim2.5\times10^{-3}$ and
$({m^2_{\tau}\over M^2_W})\sim5\times10^{-4}$,
$({m^2_{\mu}\over M^2_W})\sim1.5\times10^{-6}$ and
$({m^2_{e}\over M^2_W})\sim4\times10^{-11}$,
it is clear that for the decay
$\tau\rightarrow e\nu_{\tau}\bar\nu_e$ the only contribution with the Higgs
boson exchange which can be important is the one of the second type to the
$\tau$ vertex (and this is for large values of $\tan\beta$, say,
$\tan\beta > 20$). Together with the corresponding contribution to the
renormalized $\nu_\tau$ self energy the following Higgs   contribution to
$\Delta r^{\prime}_{\tau,e}$ is obtained in the limit of large $\tan\beta$:
\begin{eqnarray}
\delta(\Delta r^{\prime}_{\tau,e}) &=&
-{g^2_2\over128\pi^2}\left({m_{\tau}\over M_W}\tan\beta\right)^2 \\
&\times&\left[-2 + I(A^0,H^\pm) + \cos^2\alpha I(H^0,H^\pm)
+ \sin^2\alpha I(h^0,H^\pm)\right]\nonumber
\end{eqnarray}
where
\begin{eqnarray}
I(1,2) = {1\over2}{m^2_1 + m^2_2\over m^2_1 - m^2_2}\log{m^2_1\over m^2_2}
\end{eqnarray}
and $\alpha$ is the mixing angle which diagonalizes the $CP$-even
Higgs bosons mass matrix. The remaining new (supersymmetric)
contribution to $\Delta r^{\prime}_{l,l^{\prime}}$ can be found in
Appendix A of ref.\cite{MY1}.

Similar corrections have  to be included
for the decay $\tau\rightarrow \mu\nu_{\tau}\bar\nu_{\mu}$,
in the limit of large $\tan\beta$. In addition, in this case,
the tree level $H^{\pm}$ exchange may become important (for a not too
heavy $H^{\pm}$) and comparable with the genuine SUSY vertex and box
corrections. The effective Fermi constant is then given by the formula (6).

For the sake of definiteness, we begin the discussion of the results with
the ratio $G_{\tau,e}/G_{\mu,e}$, which is given by eq.(9). The
corrections $\Delta r^{\prime}_{l,l^{\prime}}$ depend, in general, on
the chargino, neutralino and slepton masses and weakly (through the
Higgs exchanges in the loop) on $\tan\beta$ and $M_{H^{\pm}}$ (the
two parameters specify completely the Higgs sector). In case of no
{\sl Left $-$ Right} mixing in the slepton mass matrix the results
depend on the masses of the left$-$handed sfermions only. They can be
parameterized by the sneutrino masses $M_{\tilde\nu_I}$ (I=1,2,3) and
are given by the relation:
\begin{eqnarray}
M^2_{\tilde l_L^I} = M^2_{\tilde\nu_I} + m^2_{l_I} - M^2_W\cos2\beta
\end{eqnarray}
It is natural to study the ratio $G_{\tau,e}/G_{\mu,e}$ as a function
of the slepton masses (which can give the non$-$universal contribution),
for several different sets of values of the "universal" variables: the
chargino and neutralino masses, $\tan\beta$ and $M_{H^{\pm}}$.
Furthermore, the considered ratio depends on four SUSY vertex corrections,
one of them being common for both $G_{\tau,e}$ and $G_{\mu,e}$.
We choose to fix the mass of the sneutrino corresponding to the "common"
vertex and study the ratio $G_{\tau,e}/G_{\mu,e}$ as a function of:
\begin{eqnarray}
x_{\tau,\mu} = { M^2_{\tilde\nu_{\mu}}
- M^2_{\tilde\nu_{\tau}}
\over M^2_{\tilde\nu_{\mu}} + M^2_{\tilde\nu_{\tau}}}
\end{eqnarray}
The results are are shown in Figs.1 and 2 where we plot the ratio
$G_{\tau,e}/G_{\mu,e}$ as a function of $x_{\tau,\mu}$ for four
different masses of the "same" $\tilde\nu_e$ sneutrino. Each point
in the shadowed area correspond to a pair of values
$(M_{\tilde\nu_{\tau}},M_{\tilde\nu_e})$ (which we scanned in
the range 50$-$500 GeV) demonstrating the (weak) dependence on the variable
complementary to $x_{\tau,\mu}$.

The pattern of those results can be understood as follows.
For low value of the "same" sneutrino mass $M_{\tilde\nu_e}$ both
$\Delta r^{\prime}_{\tau,e}$ and $\Delta r^{\prime}_{\mu,e}$ in
eq.(7) receive similar corrections and
$G_{\tau,e}/G_{\mu,e}\sim1$. When both $M_{\tilde\nu_{\tau}}$ and
$M_{\tilde\nu_{\mu}}$ are heavy,
this happens because  each of $\Delta r^{\prime}$
in eq.(7) receives
contribution only from the corrections to the "same"
vertex $e W^{\pm}\nu_e$, the corrections to other vertices and boxes
being suppressed by the inverse heavy mass squared according to the
Appelquist-Carazzone decoupling theorem. If, e.g, the $\tau$ sneutrino
becomes light (positive $x_{\tau,\mu}$) then
$\Delta r^{\prime}_{\tau,e}$ receives additional negative contribution
  from the $\tau W^{\pm}\nu_{\tau}$ vertex. However, at the same time
there is also a substantial contribution from SUSY box diagram to
$\Delta r^{\prime}_{\tau,e}$ which is positive and as a result
of the cancelation \cite{MY1} one gets again
$G_{\tau,e}$ only slightly different  from $G_{\mu,e}$. On the other hand,
for a heavy "same" sneutrino mass $M_{\tilde\nu_e}$
only $\tau W^{\pm}\nu_{\tau}$ and $\mu W^{\pm}\nu_{\mu}$ vertices can
give substantial negative contributions depending on which of the
$\tau$ and $\mu$ sneutrino masses is light, thus explaining the
pattern of the plots for heavy $M_{\tilde\nu_e}$.

   For fixed $\tan\beta$ the effects vanish, of course, with increasing
chargino and neutralino masses roughly as $M_W^2/M^2_{min(C,N)}$.
This can be seen by comparing Figs. 1 and 2.
The results depend also on $\tan\beta$, and are bigger for smaller values
of $\tan\beta$, for fixed values of the physical chargino and neutralino
masses. The $\tan\beta$ dependence is such, that it is small for fixed
values of the $M_{gau}$ and $\mu$ parameters
specifying the chargino and
neutralino masses.

Exactly analogous results hold for the other two ratios
$G_{\tau,\mu}/G_{\mu,e}$ and $G_{\tau,\mu}/G_{\tau,e}$ (with obvious
interchange of $x_{\tau,\mu}$ into $x_{\tau,e}$ and $x_{\mu,e}$
respectively) for $\tan\beta<15$, i.e. as long as the tree level
Higgs exchange can be neglected.

Next we discuss the results for the ratio $G_{\tau,\mu}/G_{\mu,e}$
for large $\tan\beta$ values (and light charged Higgs boson).
In that case, as can be seen from eq.(6) and Figure 3,
the contribution of the tree level $H^{\pm}$ exchange lowers the predicted
value of the $G_{\tau,\mu}$.
The effects of the Higgs boson 1-loop contributions to the vertices
remain, however, unnoticeable.

Fig.4 shows how in the large $\tan\beta$ case
the non$-$universal effects decrease with the increasing chargino masses
(this effect is similar as for
the $G_{\tau,e}/G_{\mu,e}$) and increasing
charged Higgs boson mass (a
smaller overall shift of the whole plot toward
negative values).

The mixing between the left and right-handed charged sleptons is proportional
to the corresponding lepton mass and therefore is not expected to play a
significant role for selectrons and smuons. For staus it can be
significant and leads to a) lower values of the lighter $\tilde\tau$
slepton (for the same tau sneutrino mass), b) admixture to it of the
right$-$handed component, which is not active in the        vertex. Both
effects tend to cancel each other. In order to demonstrate how the
{\sl Left-Right} mixing in the stau mass matrix changes the results described
above we plot in Fig.5 the ratio  $G_{\tau,\mu}/G_{\mu,e}$ as a function of
the lighter stau mass $M_{\tilde\tau_1}$ for
$M_{\tilde\nu_{\mu}}=M_{\tilde\nu_e}=500$ GeV.
The numbers in the parentheses denote the masses of $\tilde\nu_{\tau}$ and
$\tilde\tau_R$. At the right-most point
of each curve
the off-diagonal entry of the stau mass matrix vanishes.
At those points
 the mass of $\tilde\tau_1$ (which for the values of the parameters shown
in Fig.5 corresponds then
to $\tilde\tau_L$) is given                 by eq.(13).
The {\sl Left-Right} mixing increases from the right to the
left of the plot until the lighter $\tilde\tau_1$ mass reaches 45 GeV -
the current experimental limit. As can be seen, with
the {\sl Left-Right} mixing  the deviation of the ratio
$G_{\tau,\mu}/G_{\mu,e}$  from unity slightly increases
as compared to the case with no {\sl Left-Right} mixing and
the same value of $M_{\tilde\nu_{\tau}}$. Note also that with the
{\sl Left-Right} mixing the same value of $M_{\tilde\tau_1}$ may
correspond to different predictions for $G_{\tau,\mu}/G_{\mu,e}$
(of course for different $M_{\tilde\nu_{\tau}}$).

For the future  precision of $G_{\tau,\mu}$ and $G_{\tau,e}$
measurements
\cite{EXP,hobbs,cleo,EXP2} of
order 0.1\% ($G_{\mu,e}$ is known with 0.002\% precision) the only
effect that eventually can be observed is the slightly smaller
value of $G_{\tau,\mu}$ as compared to $G_{\tau,e}$ and $G_{\mu,e}$
(which, within this accuracy, should coincide in MSSM). If measured, such
effect would mean a rather precise information about MSSM: large
$\tan\beta > 20$, small $M_{H^{\pm}}\sim M_W$ (corresponding to small
$M_{A^0}$), light charginos and neutralinos and large hierarchy in the slepton
masses: $M_{\tilde\nu_{\tau}}\ll M_{\tilde\nu_{\mu}}\sim M_{\tilde\nu_e}$
This last point is in qualitative agreement with the tendency
observed in the RGE
evolution of the soft SUSY breaking terms from $M_{Planck}$ down to
$M_W$: the large (in the case of large $\tan\beta$) Yukawa coupling of
the tau drives its sneutrino mass to lower values than the masses of
$\tilde\nu_{\mu}$ and $\tilde\nu_e$.

We remark also that the new evidence from CDF for the heavy top
quark, $M_t=(174\pm 20)$ GeV,   is consistent with the minimal
$SO(10)$ Yukawa coupling unification $Y_t=Y_b$ and very large
$\tan\beta$ values, $\tan\beta =50-60$ \cite{tan}.
For such values of $\tan\beta$, the tree level charge Higgs
exchange effect seen in Fig.3 and Fig.4 is amplified by
factor 4-5. The accuracy  0.1\% for $G_{\tau,\mu}$ will be
sufficient to confirm or rule out the large $\tan\beta$
scenario with $M_{H^+}=(100-200)$ GeV.

If the precision of the $G_{\tau,\mu}$ and $G_{\tau,e}$
measurements reaches 0.01\% accuracy then, as can be seen from our
results, the sparticle (or in particular the slepton) masses become
strongly constrained.

Finally we comment on the non$-$universal contributions to the decays
$\pi\rightarrow\mu\nu_{\mu}$ and $\pi\rightarrow e\nu_e$. For
heavy enough squarks, say $M_{\tilde q}>300$ GeV, our results obtained
for  $G_{\tau,\mu}/G_{\tau,e}$ with $M_{\tilde\nu_{\tau}}>300$ GeV
apply directly to the pion decays, too. Indeed, in this case the only
non$-$negligible source of non$-$universality in both $\pi$ and
$\tau$ decays can be the supersymmetric contributions to the $\mu$
and $e$ vertices.  The present experimental value
$(g_{\mu}/g_e)_{\pi} = 1.0014\pm0.0016$ \cite{EXP2} is, however, not
accurate enough
to draw any firm conclusion (apart from the statement that $M_{\tilde
e}$ cannot be much heavier than $M_{\tilde\mu}$). For lighter squarks
a complete perturbative calculation of the effective four$-$Fermi
lagrangian is necessary. As usual, at any given order, the results will
depend on the cut$-$off separating the perturbative from
non$-$perturbative regimes.


\newpage
{\bf FIGURE CAPTIONS}
\vskip 0.5cm

\noindent {\bf Figure 1.} Departure from universality in the
$\tau\rightarrow e\nu_{\tau}\bar\nu_e$  decay as a function of
the variable $x_{\tau,\mu}$ defined in the text, for four different
values of the $\tilde\nu_e$ mass.
\vskip 0.4cm

\noindent {\bf Figure 2.} The same as in Figure 1 but for smaller
chargino and neutralino masses.
\vskip 0.4cm

\noindent {\bf Figure 3.} Departure from universality in
$\tau\rightarrow \mu\nu_{\tau}\bar\nu_{\mu}$  decay as a function of
the variable $x_{\tau,e}$ defined in the text, for four different
values of the $\tilde\nu_{\mu}$ mass.
\vskip 0.4cm

\noindent {\bf Figure 4.} The same as in Figure 3 but for different
charged Higgs boson mass $M_{H^{\pm}}$ and larger chargino and
neutralino masses.
\vskip 0.4cm

\noindent {\bf Figure 5.} Effects of the {\sl Left-Right} mixing in the
$\tilde\tau_1,\tilde\tau_2$ mass matrix. The departure from universality
is shown here as a function of the lighter $\tilde\tau$ mass.
$M_{\tilde\nu_e}=M_{\tilde\nu_{\mu}}=500$ GeV and the numbers in the
parentheses correspond to the masses (in GeVs) $(M_{\tilde\nu_{\tau}},
M_{\tilde\tau_R})$. At the rightmost points of each curve       the
mixing vanishes.
\vskip 0.4cm

\end{document}